\documentclass[a4paper]{jpconf}
\usepackage{graphicx}

\newcommand{\ttbar}{\ensuremath{t\bar{t} }\;} 

\newcommand{\tttt}{\ensuremath{t\bar{t}t\bar{t}}\;} 

\newcommand{\ttbb}{\ensuremath{t\bar{t}b\bar{b}}\;}

\begin{document}
\title{Search for standard model production of four top quarks in the lepton + jets channel in pp collisions at $\sqrt{s} = 8 $ TeV}

\author{James Keaveney on behalf of the CMS Collaboration}

\address{Pegasus Marie Curie fellow at Vrije Universiteit Brussels}

\ead{james.keaveney@vub.ac.be}

\begin{abstract}
A search is presented for standard model (SM) production of four top quarks ($t\bar{t}t\bar{t}$) in pp collisions in the lepton + jets channel. The data correspond to an integrated luminosity of 19.6 fb$^{-1}$ recorded at a centre-of-mass energy of 8 TeV with the CMS detector at the CERN LHC. A combination of kinematic reconstruction and multivariate techniques is used to distinguish between the small signal and large background. The data are consistent with expectations of the SM, and an upper limit of 32 fb is set at a 95\% confidence level on the cross section for producing four top quarks in the SM, where a limit of $32\pm{17}$ fb is expected.
\end{abstract}

\section{Introduction} 
A search is presented for standard model (SM) production of four top quarks ($t\bar{t}t\bar{t}$) in pp collisions in the lepton + jets channel performed by the CMS collaboration \cite{Khachatryan:2014sca}. The cross section for SM $t\bar{t}t\bar{t}$) production at the LHC is predicted, at leading order, to be extremely small: $\sigma^{SM}_{t\bar{t}t\bar{t}} \approx$ 1 fb at $\sqrt{s} = 8$ TeV \cite{threeTop}.

In many models beyond the SM (BSM) involving supersymmetry, massive coloured bosons, Higgs boson or top quark compositeness, or extra dimensions, $\sigma_{t\bar{t}t\bar{t}}$ is enhanced \cite{ttttNLO,AguilarSaavedra:2011ck,Plehn:1173428,GoncalvesNetto:2012nt,Fuks:2012im,Calvet:2012rk,Dev:2014yca}. In certain regions of BSM parameter space, these final states have kinematics similar to those of SM $t\bar{t}t\bar{t}$ production. In such cases, reinterpretation of an upper limit on SM production of $t\bar{t}t\bar{t}$ has the potential to constrain BSM theories. Moreover, in direct searches for these BSM signatures, SM production of $t\bar{t}t\bar{t}$ can be a background. Hence, experimental constraints on $\sigma_{t\bar{t}t\bar{t}}$ have the potential to enhance the discovery reach of such searches.

\section{Event selection}
The selections require the presence of one well-identified and isolated muon or electron with $P_{T} >$ 30 GeV and with respective muon or electron $|\eta| <$ 2.1 or 2.5. Jets are required to have $P_{T} >$  30 GeV and $|\eta| <$ 2.5. All events are required to have the number of selected jets ($N_{jets}$) to be at least six. Events are required to have the number of jets identified as originating from a b quark ($N_{btags}$) to be at least two. The $H_{T}$ of an event is defined as the scalar sum of the $P_{T}$ of all the selected jets in the event. Events are also required to have $H_{T} >$ 400 GeV and $E_{T}^{miss} >$ 30 GeV. 

\section{Event classification with an MVA algorithm\label{sec:bdt}}
In the lepton + jets channel, the presence of three jet-decaying top quarks in \tttt events can be exploited to distinguish such events from \ttbar background events, which contain only one jet-decaying top quark. The challenge in the kinematic reconstruction of such top quarks in an event containing many jets is to find correct selections of three jets that arise from any single top quark when many incorrect three-jet combinations are possible. Such correctly selected combinations are referred to as "correct trijets", while combinations containing one or more jets not originating from the same top quark are referred to as "incorrect trijets".  
The large number of incorrect trijets in signal and background events motivates the use of MVA methods to distinguish between correct and incorrect trijets by combining information from a set of input variables. Six variables which discriminate between correct and incorrect trijets are combined in a boosted decision tree algorithm ($BDT_{trijet}$) using the {\sc TMVA} package \cite{Hocker:2007ht}.

\begin{figure}[h]
\begin{center}
\begin{minipage}{12pc}
\includegraphics[width=12pc]{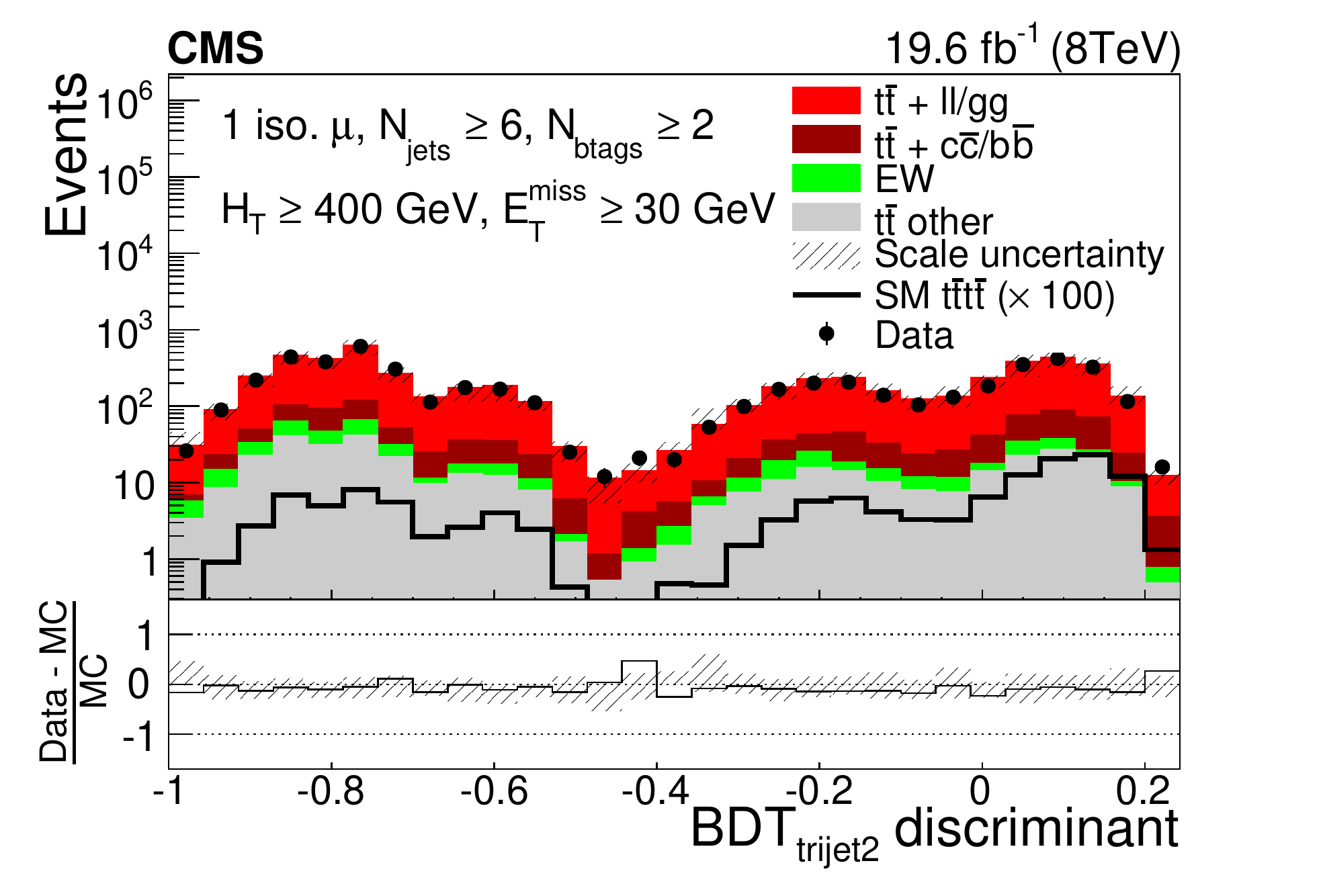}
\end{minipage}\hspace{1pc}%
\begin{minipage}{12pc}
\includegraphics[width=12pc]{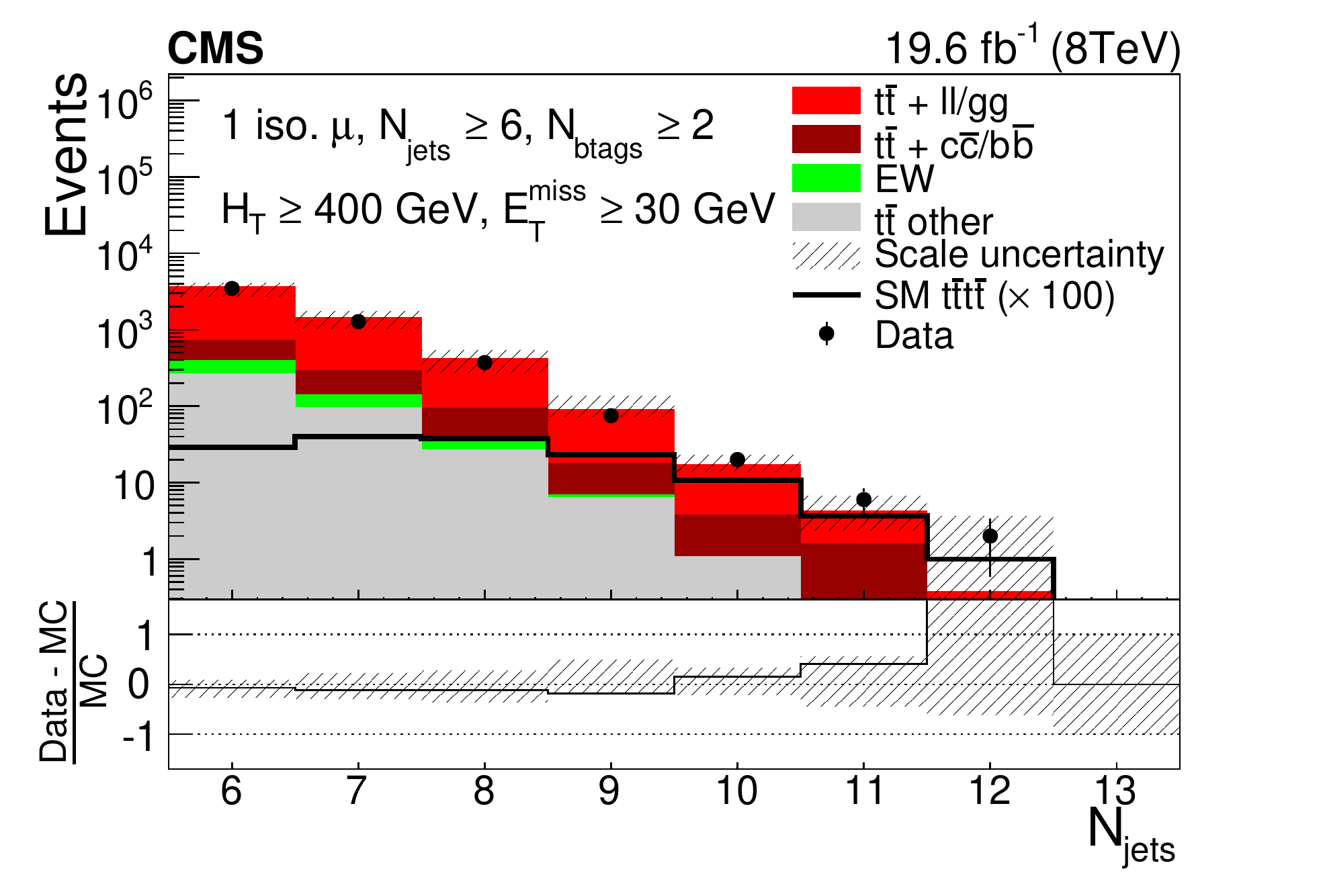}
\end{minipage} 
\begin{minipage}{12pc}
\includegraphics[width=12pc]{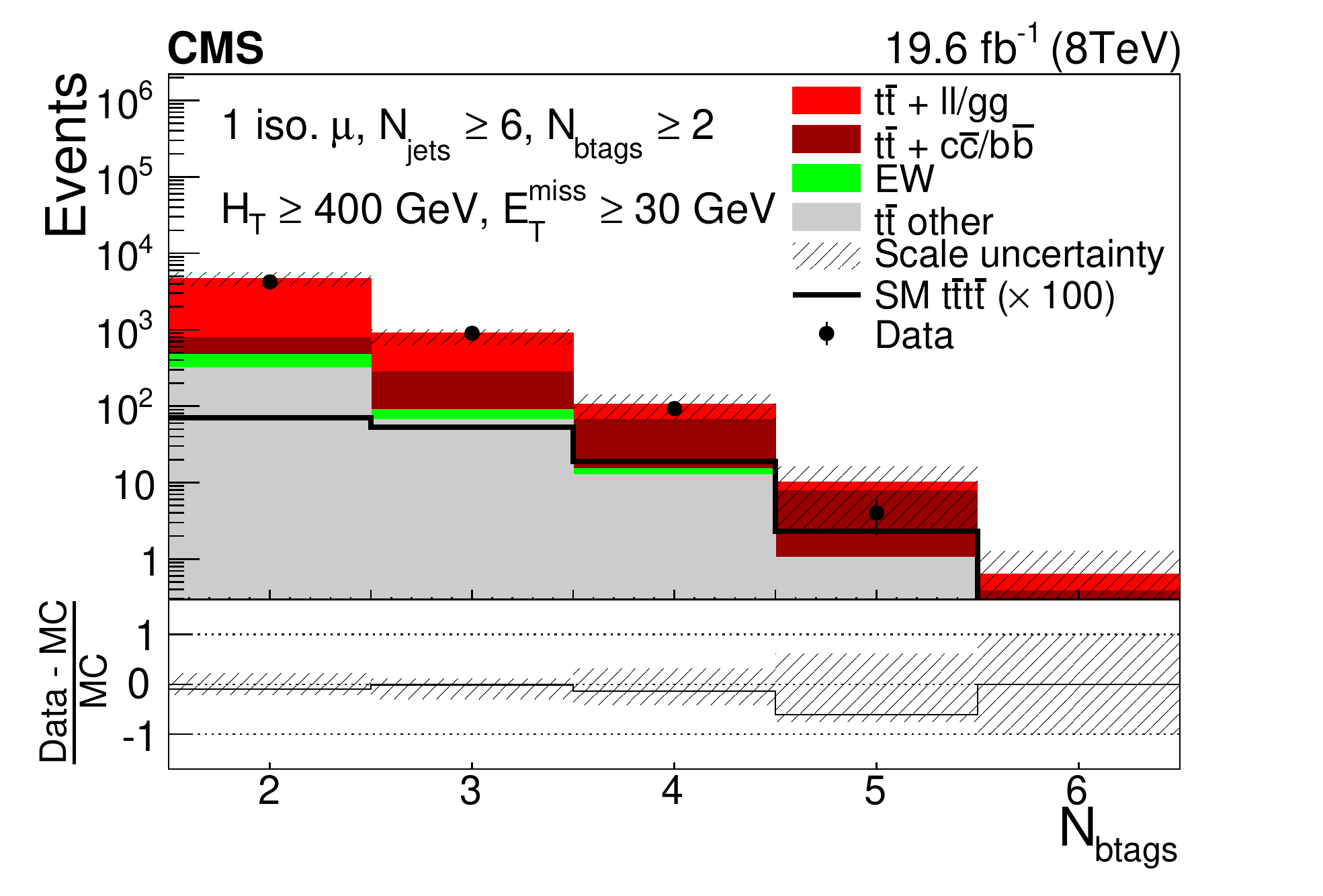}
\end{minipage} \\
\begin{minipage}{12pc}
\includegraphics[width=12pc]{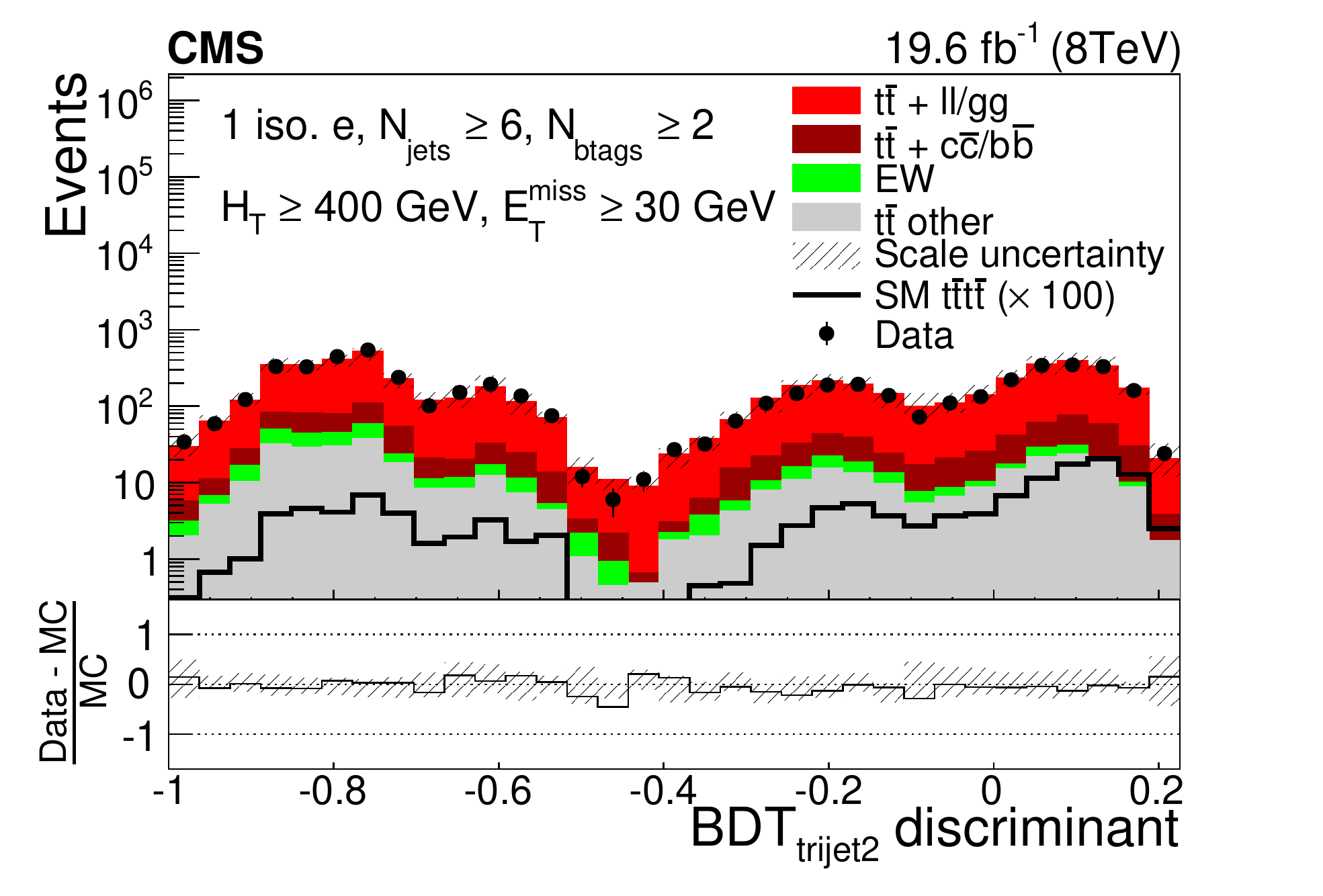}
\end{minipage}\hspace{1pc}%
\begin{minipage}{12pc}
\includegraphics[width=12pc]{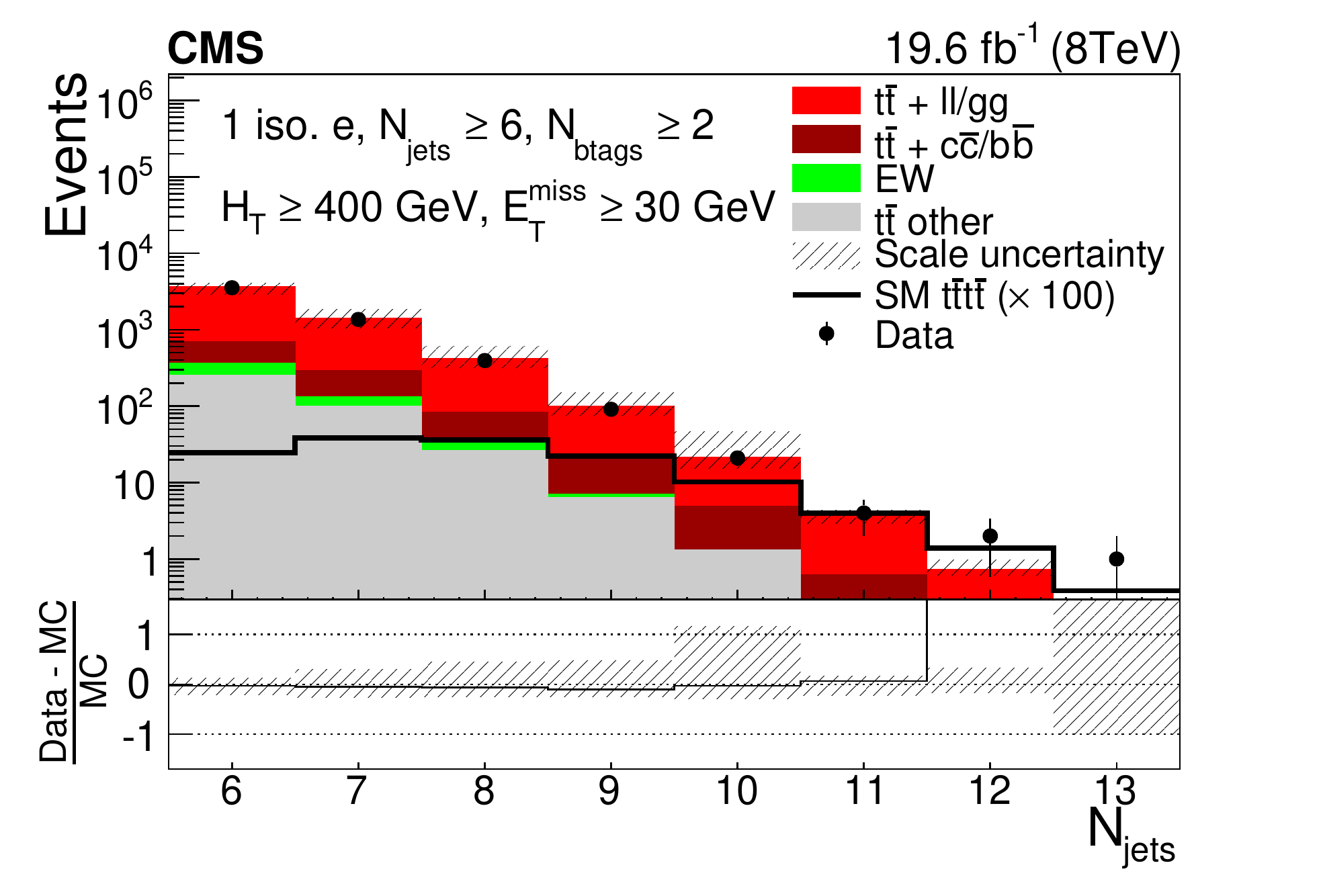}
\end{minipage} 
\begin{minipage}{12pc}
\includegraphics[width=12pc]{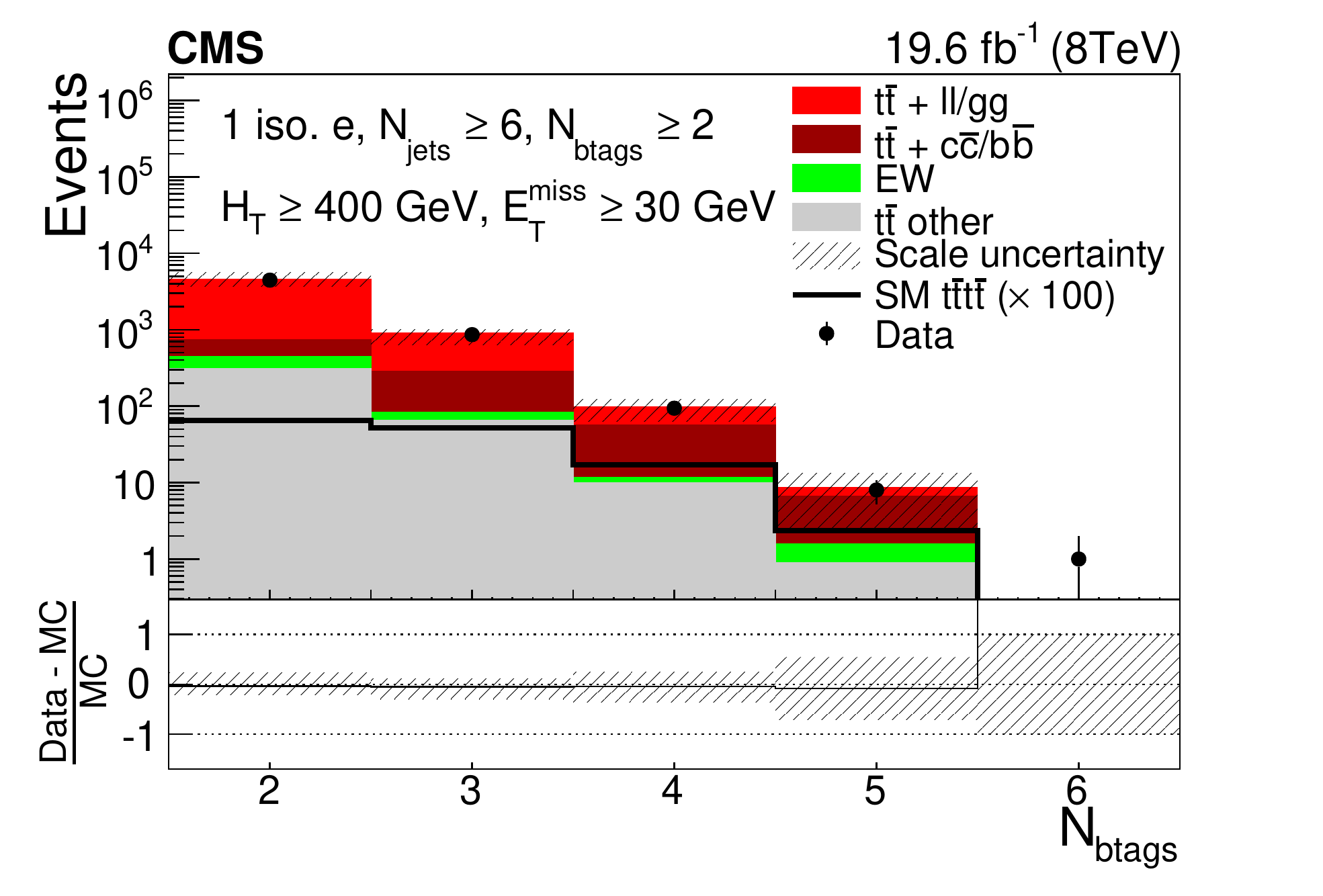}
\end{minipage}
\caption{The distribution in (left to right) the $BDT_{trijet2}$ discriminant, $N_{jets}$, and in $N_{btags}$ for the $\mu$ + jets (top row) and e + jets channels (bottom row). The ratios plotted at the bottom of each panel reflect the percent differences between data and MC events. The hatched areas show the changes in the calculated predictions produced by factors of two and one half changes in the factorisation and renormalisation scales in the \ttbar simulation.}
\label{fig:BCvar}
\end{center}
\end{figure}

The trijet with the largest value of $BDT_{trijet}$ discriminant is removed from the event, and the $BDT_{trijet}$ discriminant of the trijet of the remaining jets with highest value ($BDT_{trijet2}$) is used to distinguish between \tttt and \ttbar events. Distributions in this variable, in $N_{jets}$ and in $N_{btags}$ in data and simulation are shown in Fig. \ref{fig:BCvar}. The reduced event (RE) is constructed by subtracting the jets contained in the highest $BDT_{trijet}$ ranking trijet. Two variables based on the RE are (i) $H^{RE}_{T}$, which is the $H_{T}$ of the RE and (ii) $M^{RE}$, which is  the invariant mass of the system comprising all the jets in the RE. Because \tttt events can contain up to ten hard jets from top quark decays, while \ttbar events contain up to four, the following variables based on jet activity of the event possess discrimination power: (i) $N_{jets}$, (ii) $H^{b}_{T}$, (iii) $H_{T}/H_{p}$, (iv) $H^{ratio}_{T}$, (v) $p_{T5}$, and (vi) $p_{T6}$. The $H^{b}_{T}$ variable is the $H_{T}$ of the b-tagged jets. In the $H_{T}/H_{p}$ ratio, $H_{p}$ is the scalar sum of the total momenta of the selected jets. The ratio of the $H_{T}$ of the four leading jets to the $H_{T}$ of the other jets is defined as $H^{ratio}_{T}$. The $p_{T5}$ and $p_{T6}$ variables represent, respectively, the $P_{T}$ values of jets of 5th and 6th largest $P_{T}$. The multiplicity of b-tagged jets is also used in the discriminant.

\begin{figure}[h]
\begin{center}
\begin{minipage}{12pc}
\includegraphics[width=12pc]{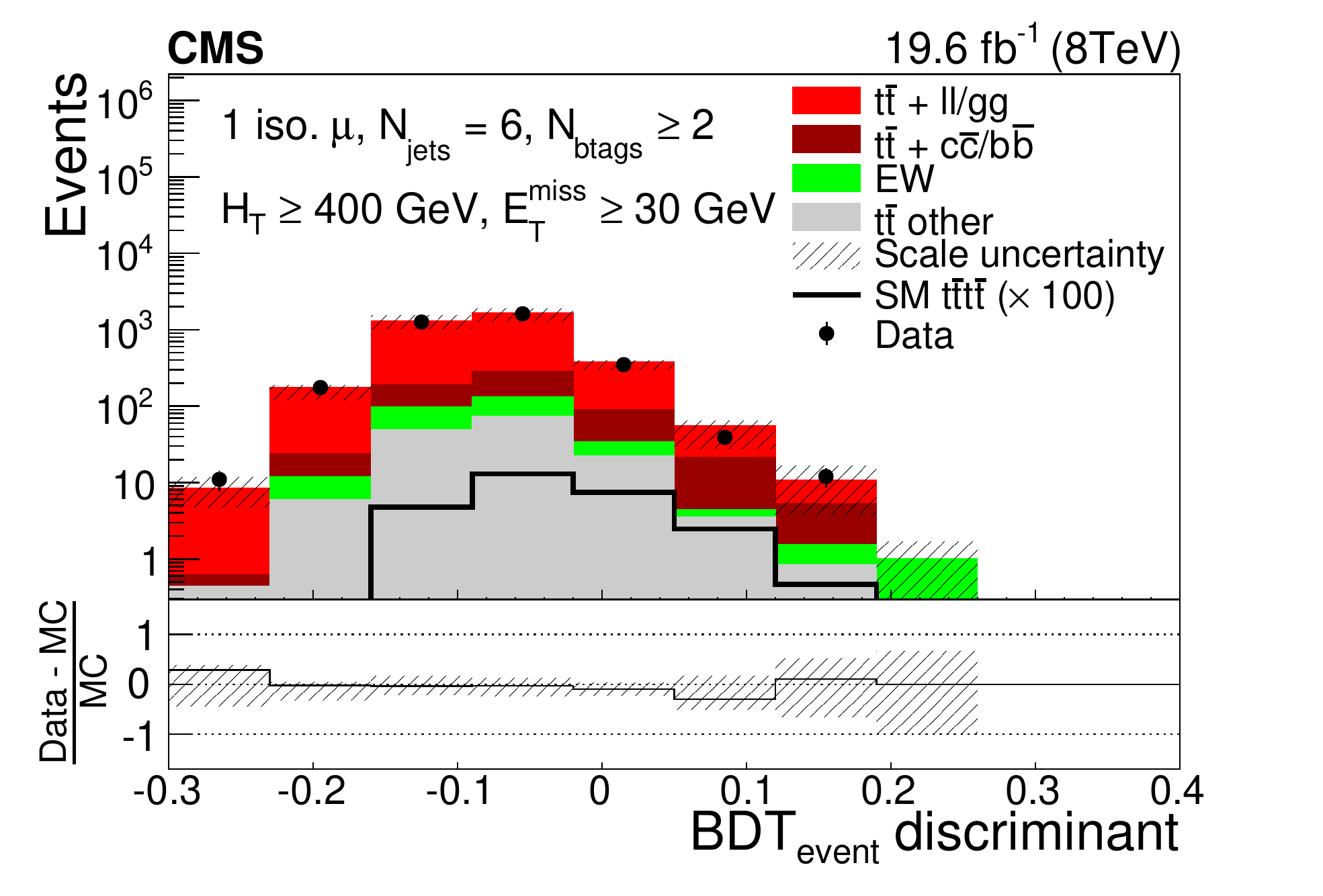}
\end{minipage}\hspace{1pc}%
\begin{minipage}{12pc}
\includegraphics[width=12pc]{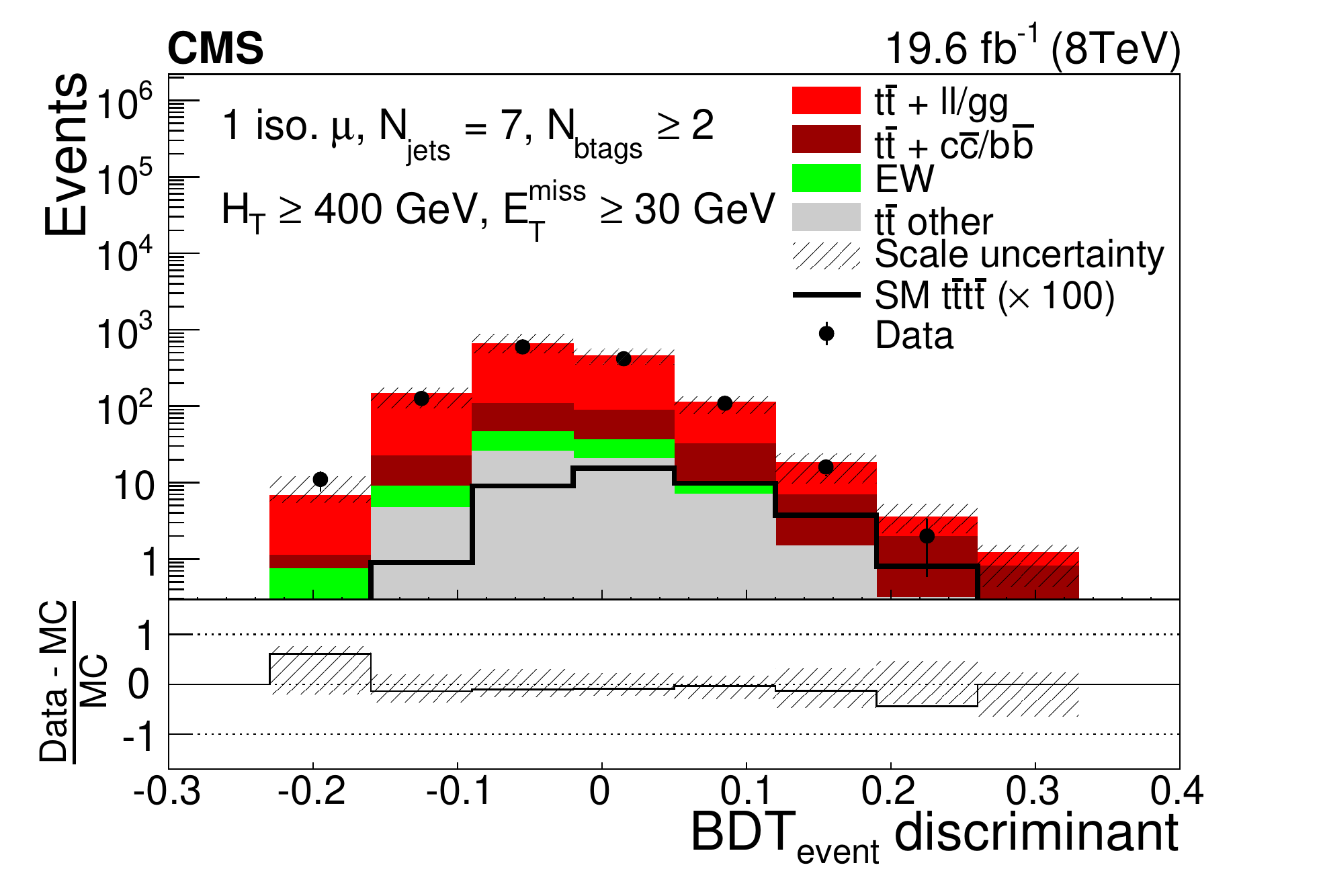}
\end{minipage} 
\begin{minipage}{12pc}
\includegraphics[width=12pc]{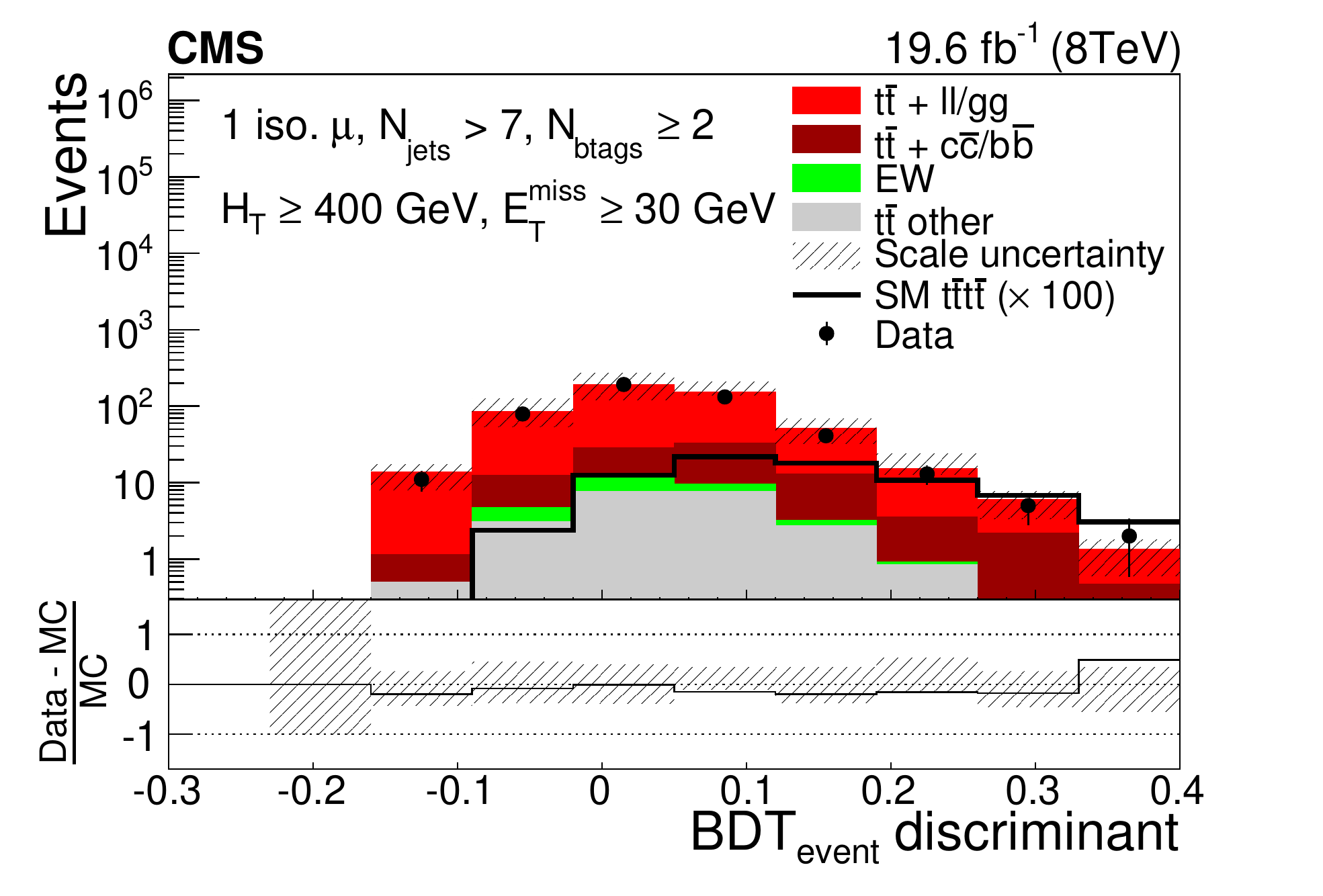}
\end{minipage} \\
\begin{minipage}{12pc}
\includegraphics[width=12pc]{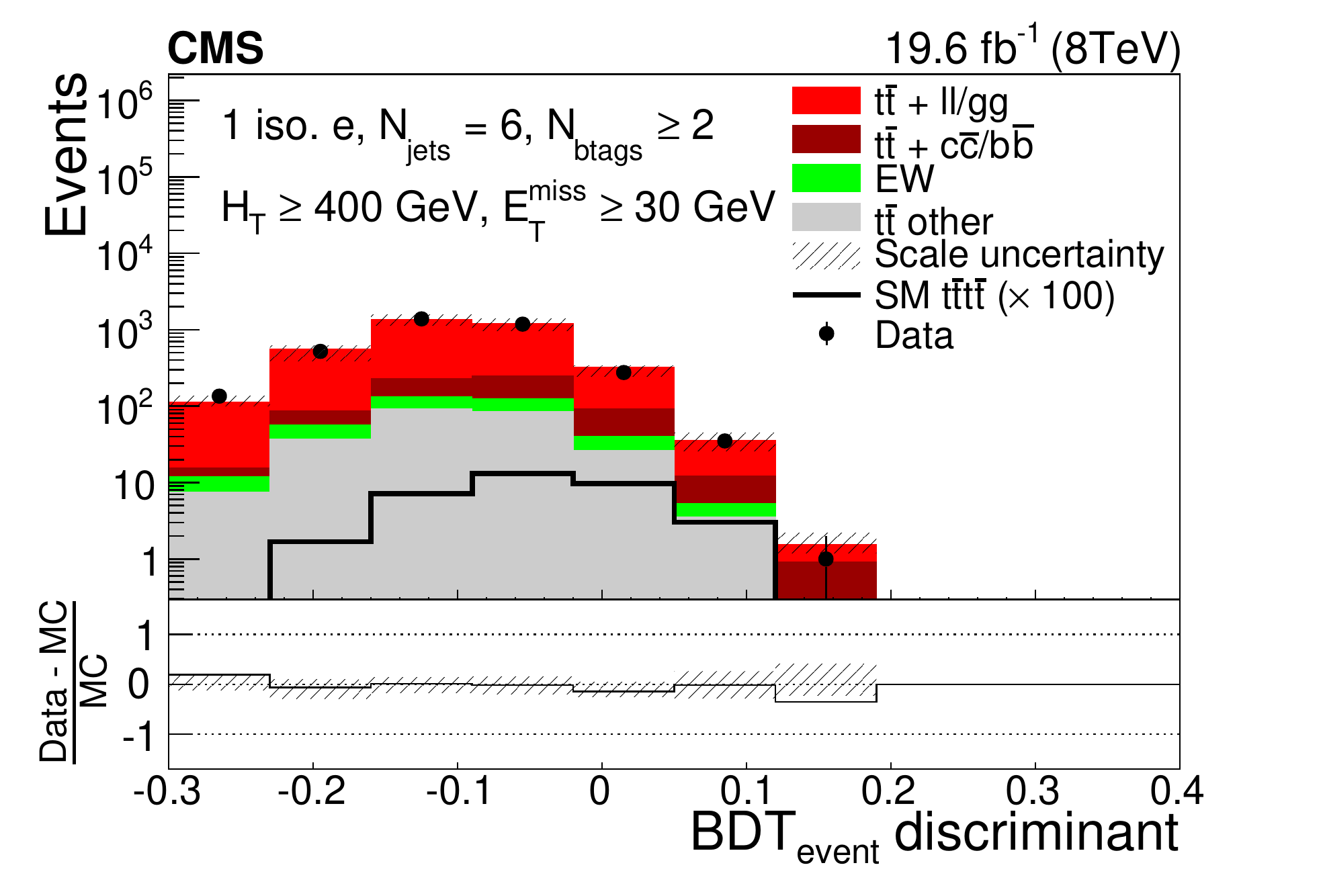}
\end{minipage}\hspace{1pc}%
\begin{minipage}{12pc}
\includegraphics[width=12pc]{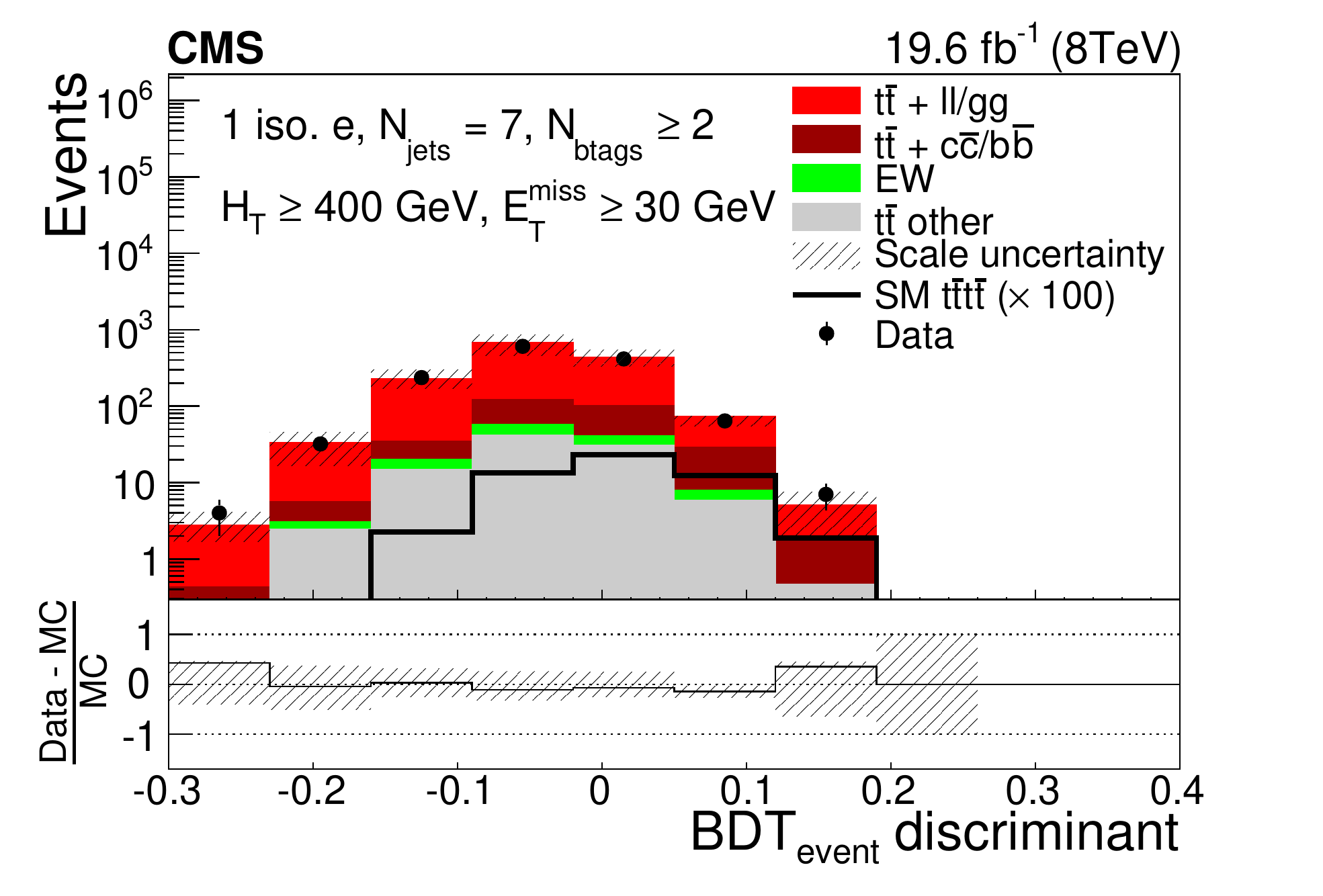}
\end{minipage} 
\begin{minipage}{12pc}
\includegraphics[width=12pc]{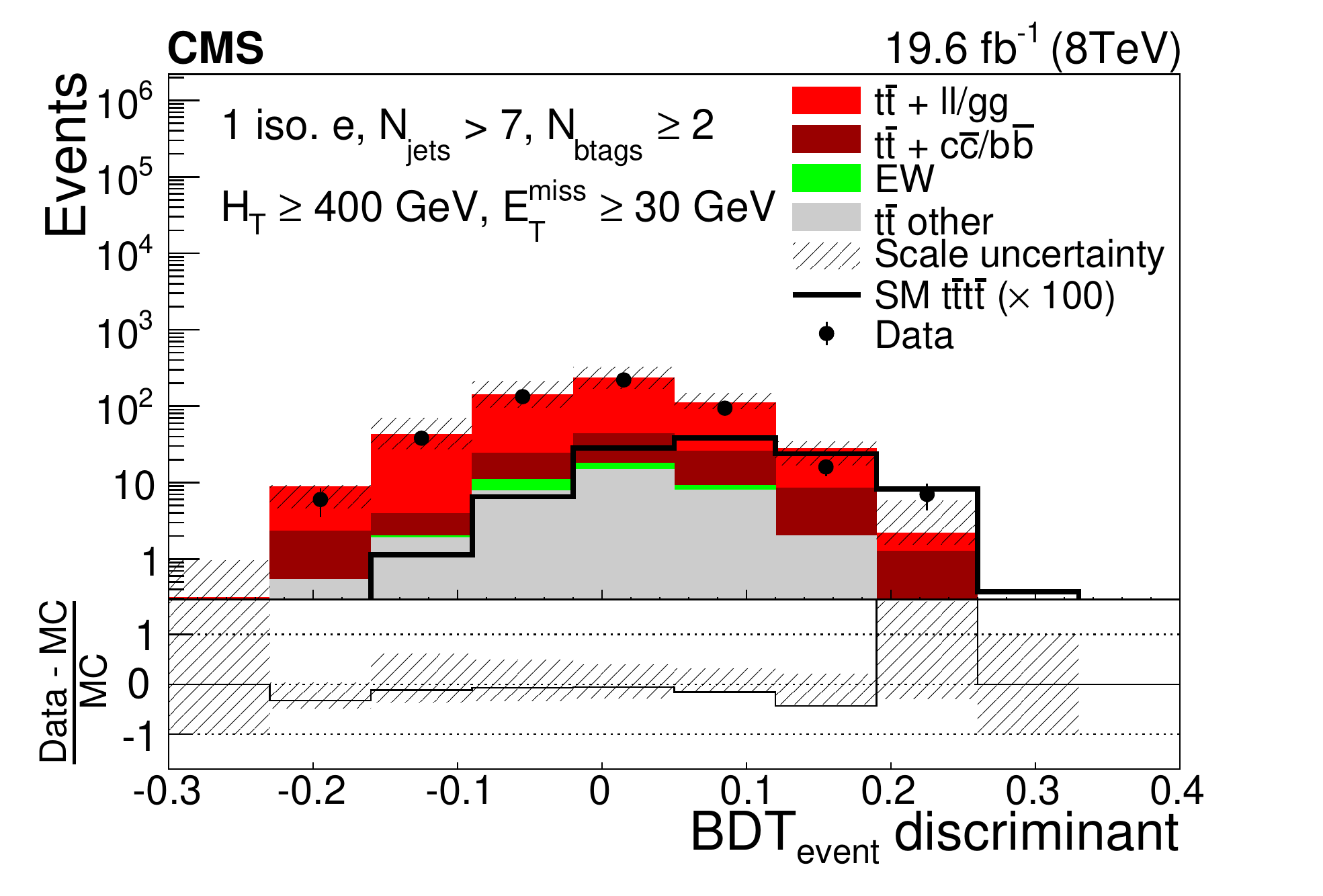}
\end{minipage}
\caption{The distribution in the $BDT_{event}$ discriminant for data and simulation in events with (left to right) $N_\mathrm{jets} = 6$, $N_\mathrm{jets} = 7$ and $N_\mathrm{jets} > 7$ for the $\mu$ + jets (top row) and e + jets channels (bottom row). The ratios plotted at the bottom of each panel reflect the percent differences between data and MC events. The hatched areas reflect the changes in the calculated predictions produced by factors of two and one half changes in the factorisation and renormalisation scales.}
\label{fig:BCevent}
\end{center}
\end{figure}

The ten variables described are combined using a second, event-level  BDT ($BDT_{event}$). To maximise sensitivity, the events are divided into three categories corresponding to $N_\mathrm{jets}$ = 6, 7 and, $>$ 7. In Fig. \ref{fig:BCevent}, distributions in the $BDT_{event}$ discriminant are shown in data and in MC for each of these categories.

\section{Systematic uncertainties and limit setting}
The systematic uncertainties considered in this analysis are separated into two categories: (i) those that affect the normalisations of the $BDT_{event}$ discriminant distributions of both signal and backgrounds and (ii) those that affect the form of the distributions of just the backgrounds. The normalisations are affected by the uncertainty in integrated luminosity of the data and the theoretical cross sections of the signal and background processes. An uncertainty in the integrated luminosity of 2.6\% is included \cite{CMS-PAS-LUM-13-001}. The uncertainty in the \ttbar cross section is expected to dominate, and is taken from Ref. \cite{PhysRevLett.110.252004} as $^{+ 2.5\%}_{ - 3.4\%}$~(factorisation and renormalisation scales) and $^{+ 2.5\%}_{ - 2.6\%}$~(PDF). A range of theoretical and experimental sources of systematic uncertainty can affect the form of the distributions of the $BDT_{event}$ discriminant. As \ttbar is the dominant background, systematic effects on the form of the distributions are considered only for \ttbar events. The impact of contributions from higher-order corrections in the \ttbar simulation is quantified by comparing alternative \ttbar samples that are generated with the renormalisation and factorisation scales simultaneously changed up and down by a factor of two relative to the nominal \ttbar sample. The matching of partons originating from the matrix element to the jets from the parton showers is performed according to the MLM prescription \cite{Mangano:2006rw}. The uncertainty arising from this prescription is estimated by changing the minimum $k_t$ measure between partons by factors of 0.5 and 2.0 and the jet matching threshold by factors of 0.75 and 1.5. To evaluate the uncertainty due to imperfect knowledge of the JES, JER, b tagging, and lepton-identification efficiencies, and the cross section for minimum-bias production used in the pileup-reweighting procedure in simulation, the input value of each parameter is changed by $\pm 1$ standard deviation of its uncertainty. A systematic uncertainty due to the imperfect knowledge of the contribution from the \ttbb component in \ttbar events is also estimated.

No significant excess of events to represent SM \tttt production is observed above the background prediction. Therefore, an upper limit on $\sigma_{\tttt}$ is set by performing a simultaneous maximum likelihood fit to the distributions in the $BDT_{event}$ discriminant for signal and background in the six event categories described in Section \ref{sec:bdt}. The modified frequentist $CL_{s}$ approach \cite{Junk:1999kv,Read:2002hq} using the asymptotic approximation is adopted to measure the upper limit using the {\sc RooStats} package \cite{Cowan:2010js,RooStats}. The limit calculated at a 95\% confidence level (CL) on the production cross section $\sigma_{\tttt}$ is $32$ fb, where a limit of $32\pm{17}$ fb is expected. These limits are approximately 25 $\times$ $\sigma^{SM}_{\tttt}$.

\section{Conclusions}
A search for events containing four top quarks was performed using data collected with the CMS detector in lepton + jets final states at $\sqrt{s} = 8$ TeV, corresponding to an integrated luminosity of 19.6 fb$^{-1}$. A simultaneous maximum likelihood fit to $BDT_{event}$ discriminant distributions was performed, from which an upper limit on $\sigma_{\tttt}$ of $32$ fb was calculated at a 95\% CL, where a limit of $32\pm{17}$ fb was expected. This result has the potential to constrain BSM theories producing \tttt final states with kinematics similar to the SM process.

\end{document}